
\input epsf
\newbox\leftpage \newdimen\fullhsize \newdimen\hstitle
\newdimen\hsbody
\tolerance=1000\hfuzz=2pt
\def\printertype{ps: }
\def\qms{\def\printertype{qms: }
\ifx\answ\bigans\else\voffset=-.4truein\hoffset=.125truein\fi}
\def\bigans{b }
%
\let\answ\bigans \ifx\answ\bigans\message{This will come out
unreduced.} \magnification=1200\baselineskip=12pt plus 2pt minus 1pt
\hsbody=\hsize \hstitle=\hsize 
%
\else\message{(This will be reduced.} \let\lr=L
\magnification=1000\baselineskip=16pt plus 2pt minus 1pt
\voffset=-.31truein\vsize=7truein\hoffset=-.59truein
\hstitle=8truein\hsbody=4.75truein\fullhsize=10truein\hsize=\hsbody
\output={\ifnum\pageno=0 \shipout\vbox{\special{\printertype
landscape}\makeheadline \hbox to
\fullhsize{\hfill\pagebody\hfill}}\advancepageno \else
\almostshipout{\leftline{\vbox{\pagebody\makefootline}}}
\advancepageno \fi} \def\almostshipout#1{\if L\lr \count1=1
\message{[\the\count0.\the\count1]} \global\setbox\leftpage=#1
\global\let\lr=R \else \count1=2 \shipout\vbox{\special{\printertype
landscape} \hbox to\fullhsize{\box\leftpage\hfil#1}}
\global\let\lr=L\fi} \fi
\catcode`\@=11 
\newcount\yearltd\yearltd=\year\advance\yearltd by -1900

%
%

\def\draftmode{\message{ DRAFTMODE }\def\draftdate{{\rm preliminary
draft:
\number\month/\number\day/\number\yearltd\ \ \hourmin}}%
\headline={\hfil\draftdate}\writelabels\baselineskip=20pt plus 2pt
minus 2pt
 {\count255=\time\divide\count255 by 60
\xdef\hourmin{\number\count255}
  \multiply\count255 by-60\advance\count255 by\time
  \xdef\hourmin{\hourmin:\ifnum\count255<10 0\fi\the\count255}}}
\def\nolabels{\def\wrlabel##1{}\def\eqlabeL##1{}\def\reflabel##1{}}
\def\writelabels{\def\wrlabel##1{\leavevmode\vadjust{\rlap{\smash%
{\line{{\escapechar=`
\hfill\rlap{\sevenrm\hskip.03in\string##1}}}}}}}%
\def\eqlabeL##1{{\escapechar-1\rlap{\sevenrm\hskip.05in\string##1}}}%
\def\reflabel##1{%
\noexpand\llap{\noexpand\sevenrm\string\string\string##1}}}
\nolabels
%
\global\newcount\secno \global\secno=0
\global\newcount\meqno \global\meqno=1
\def\newsec#1{\global\advance\secno by1\message{(\the\secno. #1)}
\global\subsecno=0\xdef\secsym{\the\secno.}\global\meqno=1
\noindent{\bf\the\secno. #1}\writetoca{{\secsym} {#1}}
\par\nobreak\medskip\nobreak}
\xdef\secsym{}
\global\newcount\subsecno \global\subsecno=0
\def\subsec#1{\global\advance\subsecno
by1\message{(\secsym\the\subsecno. #1)}
\bigbreak\noindent{\it\secsym\the\subsecno.
#1}\writetoca{\string\quad
{\secsym\the\subsecno.} {#1}}\par\nobreak\medskip\nobreak}
\def\appendix#1#2{%
\global\meqno=1\global\subsecno=0\xdef\secsym{\hbox{#1.}}
\bigbreak\bigskip\noindent{\bf Appendix #1. #2}\message{(#1. #2)}
\writetoca{Appendix {#1.} {#2}}\par\nobreak\medskip\nobreak}
%
%
\def\eqnn#1{\xdef #1{(\secsym\the\meqno)}\writedef{#1\leftbracket#1}%
\global\advance\meqno by1\wrlabel#1}
\def\eqna#1{\xdef #1##1{\hbox{$(\secsym\the\meqno##1)$}}
\writedef{#1\numbersign1\leftbracket#1{\numbersign1}}%
\global\advance\meqno by1\wrlabel{#1$\{\}$}}
\def\eqn#1#2{\xdef
#1{(\secsym\the\meqno)}\writedef{#1\leftbracket#1}%
\global\advance\meqno by1$$#2\eqno#1\eqlabeL#1$$}
%
\newskip\footskip\footskip10pt plus 1pt minus 1pt 
\def\f@@t{\baselineskip\footskip\bgroup\aftergroup\@foot\let\next}
\setbox\strutbox=\hbox{\vrule height9.5pt depth4.5pt width0pt}
\global\newcount\ftno \global\ftno=0
\def\foot{\global\advance\ftno by1\footnote{$^{\the\ftno}$}}
%
\newwrite\ftfile
\def\footend{\def\foot{\global\advance\ftno by1\chardef\wfile=\ftfile
$^{\the\ftno}$\ifnum\ftno=1\immediate\openout\ftfile=foots.tmp\fi%
\immediate\write\ftfile{\noexpand\smallskip%
\noexpand\item{f\the\ftno:\ }\pctsign}\findarg}%
\def\footatend{\vfill\eject\immediate\closeout\ftfile{\parindent=20pt
\centerline{\bf Footnotes}\nobreak\bigskip\input foots.tmp }}}
\def\footatend{}
%
%
\global\newcount\refno \global\refno=1
\newwrite\rfile
\def\ref{$^{\the\refno}$\nref}
\def\nref#1{\xdef#1{$^{\the\refno}$}\xdef\rfn{\the\refno}
\writedef{#1\leftbracket#1}%
\ifnum\refno=1\immediate\openout\rfile=refs.tmp\fi%
\global\advance\refno by1\chardef\wfile=\rfile\immediate%
\write\rfile{\noexpand\item{{\rfn}.\
}\reflabel{#1\hskip.31in}\pctsign}\findarg}%
\def\findarg#1#{\begingroup\obeylines\newlinechar=`\^^M\pass@rg}%
{\obeylines\gdef\pass@rg#1{\writ@line\relax #1^^M\hbox{}^^M}%
\gdef\writ@line#1^^M{\expandafter\toks0\expandafter{\striprel@x #1}%
\edef\next{%
\the\toks0}\ifx\next\em@rk\let\next=\endgroup\else\ifx\next\empty%
\else\immediate\write\wfile{%
\the\toks0}\fi\let\next=\writ@line\fi\next\relax}}%
\def\striprel@x#1{} \def\em@rk{\hbox{}}

\def\addref#1{\immediate\write\rfile{\noexpand\item{}#1}} 
%

%
\def\startrefs#1{\immediate\openout\rfile=refs.tmp\refno=#1}
\def\xref{\expandafter\xr@f}\def\xr@f[#1]{#1}
\def\refs#1{[\r@fs #1{\hbox{}}]}
\def\r@fs#1{\edef\next{#1}\ifx\next\em@rk\def\next{}\else
\ifx\next#1\xref #1\else#1\fi\let\next=\r@fs\fi\next}
%

%
\newwrite\ffile\global\newcount\figno \global\figno=1
\def\fig{fig.~\the\figno\nfig}
\def\nfig#1{\xdef#1{fig.~\the\figno}%
\writedef{#1\leftbracket fig.\noexpand~\the\figno}%
\ifnum\figno=1\immediate\openout\ffile=figs.tmp%
\fi\chardef\wfile=\ffile%
\immediate\write\ffile{\noexpand\medskip\noexpand\item{Fig.\
\the\figno. }
\reflabel{#1\hskip.55in}\pctsign}\global\advance\figno by1\findarg}
\def\xfig{\expandafter\xf@g}\def\xf@g fig.\penalty\@M\ {}
\def\figs#1{figs.~\f@gs #1{\hbox{}}}
\def\f@gs#1{\edef\next{#1}\ifx\next\em@rk\def\next{}\else
\ifx\next#1\xfig #1\else#1\fi\let\next=\f@gs\fi\next}
\newwrite\lfile
{\escapechar-1\xdef\pctsign{\string\%}\xdef\leftbracket{\string\{}
\xdef\rightbracket{\string\}}\xdef\numbersign{\string\#}}

\def\writestop{%
\def\writestoppt{\immediate\write\lfile{\string\pageno%
\the\pageno\string\startrefs\leftbracket\the\refno\rightbracket%
\string\def\string\secsym\leftbracket\secsym\rightbracket%
\string\secno\the\secno\string\meqno\the\meqno}%
\immediate\closeout\lfile}}
\def\writestoppt{}\def\writedef#1{}
\def\seclab#1{\xdef
#1{\the\secno}\writedef{#1\leftbracket#1}\wrlabel{#1=#1}}
\def\subseclab#1{\xdef #1{\secsym\the\subsecno}%
\writedef{#1\leftbracket#1}\wrlabel{#1=#1}}
\newwrite\tfile \def\writetoca#1{}
\def\leaderfill{\leaders\hbox to 1em{\hss.\hss}\hfill}
\def\writetoc{\immediate\openout\tfile=toc.tmp
   \def\writetoca##1{{\edef\next{\write\tfile{\noindent ##1
   \string\leaderfill {\noexpand\number\pageno} \par}}\next}}}

%
\ifx\answ\bigans
 
scaled\magstep3
 \font\titlei=cmmi10
scaled\magstep3
\font\titleis=cmmi7 scaled\magstep3 \font\titleiss=cmmi5
scaled\magstep3
\font\titlesy=cmsy10 scaled\magstep3 \font\titlesys=cmsy7
scaled\magstep3
\font\titlesyss=cmsy5 scaled\magstep3 
scaled\magstep3
\else
 
scaled\magstep4
 \font\titlei=cmmi10
scaled\magstep4
\font\titleis=cmmi7 scaled\magstep4 \font\titleiss=cmmi5
scaled\magstep4
\font\titlesy=cmsy10 scaled\magstep4 \font\titlesys=cmsy7
scaled\magstep4
\font\titlesyss=cmsy5 scaled\magstep4 
scaled\magstep4
\font\absrm=cmr10 scaled\magstep1 \font\absrms=cmr7 scaled\magstep1
\font\absrmss=cmr5 scaled\magstep1 \font\absi=cmmi10 scaled\magstep1
\font\absis=cmmi7 scaled\magstep1 \font\absiss=cmmi5 scaled\magstep1
\font\abssy=cmsy10 scaled\magstep1 \font\abssys=cmsy7 scaled\magstep1
\font\abssyss=cmsy5 scaled\magstep1 \font\absbf=cmbx10
scaled\magstep1
\skewchar\absi='177 \skewchar\absis='177 \skewchar\absiss='177
\skewchar\abssy='60 \skewchar\abssys='60 \skewchar\abssyss='60
\fi
\skewchar\titlei='177 \skewchar\titleis='177 \skewchar\titleiss='177
\skewchar\titlesy='60 \skewchar\titlesys='60 \skewchar\titlesyss='60
\ifx\answ\bigans\def\abstractfont{\footfont}\else
\def\abstractfont{\def\rm{\fam0\absrm}
\textfont0=\absrm \scriptfont0=\absrms \scriptscriptfont0=\absrmss
\textfont1=\absi \scriptfont1=\absis \scriptscriptfont1=\absiss
\textfont2=\abssy \scriptfont2=\abssys \scriptscriptfont2=\abssyss
\textfont\itfam=\bigit \def\it{\fam\itfam\bigit}
\textfont\bffam=\absbf \def\bf{\fam\bffam\absbf} \rm} \fi
\def\tenpoint{\def\rm{\fam0\tenrm}
\textfont0=\tenrm \scriptfont0=\sevenrm \scriptscriptfont0=\fiverm
\textfont1=\teni  \scriptfont1=\seveni  \scriptscriptfont1=\fivei
\textfont2=\tensy \scriptfont2=\sevensy \scriptscriptfont2=\fivesy
\textfont\itfam=\tenit \def\it{\fam\itfam\tenit}
\textfont\bffam=\tenbf \def\bf{\fam\bffam\tenbf} \rm}
\def\noblackbox{\overfullrule=0pt}
\hyphenation{anom-aly anom-alies coun-ter-term coun-ter-terms}
\def\inv{^{\raise.15ex\hbox{${\scriptscriptstyle -}$}\kern-.05em 1}}

\def\Dsl{\,\raise.15ex\hbox{/}\mkern-13.5mu D} 
\def\dsl{\raise.15ex\hbox{/}\kern-.57em\partial}
\def\del{\partial}

\font\bigit=cmti10 scaled \magstep1
\def\lspace{\ifx\answ\bigans{}\else\qquad\fi}
\def\lbspace{\ifx\answ\bigans{}\else\hskip-.2in\fi} 
\def\boxeqn#1{\vcenter{\vbox{\hrule\hbox{\vrule\kern3pt\vbox{\kern3pt
	\hbox{${\displaystyle #1}$}\kern3pt}\kern3pt\vrule}\hrule}}}
\def\mbox#1#2{\vcenter{\hrule \hbox{\vrule height#2in
		\kern#1in \vrule} \hrule}}  
%

\def\darr#1{\raise1.5ex\hbox{$\leftrightarrow$}\mkern-16.5mu #1}

\def\roughly#1{%
\raise.3ex\hbox{$#1$\kern-.75em\lower1ex\hbox{$\sim$}}}
\font\tenmss=cmss10
\font\absmss=cmss10 scaled\magstep1
\newfam\mssfam
\font\footrm=cmr8  \font\footrms=cmr5
\font\footrmss=cmr5   \font\footi=cmmi8
\font\footis=cmmi5   \font\footiss=cmmi5
\font\footsy=cmsy8   \font\footsys=cmsy5
\font\footsyss=cmsy5   \font\footbf=cmbx8
\font\footmss=cmss8
\def\footfont{\def\rm{\fam0\footrm}
\textfont0=\footrm \scriptfont0=\footrms
\scriptscriptfont0=\footrmss
\textfont1=\footi \scriptfont1=\footis
\scriptscriptfont1=\footiss
\textfont2=\footsy \scriptfont2=\footsys
\scriptscriptfont2=\footsyss
\textfont\itfam=\footi \def\it{\fam\itfam\footi}
\textfont\mssfam=\footmss \def\mss{\fam\mssfam\footmss}
\textfont\bffam=\footbf \def\bf{\fam\bffam\footbf} \rm}
\catcode`\@=12 
%
\newif\ifdraft

\noblackbox
\catcode`\@=11
\newif\iffrontpage
\def\figin{\epsfcheck\figin}\def\figins{\epsfcheck\figins}
\def\epsfcheck{\ifx\epsfbox\UnDeFiNeD
\message{(NO epsf.tex, FIGURES WILL BE IGNORED)}
\gdef\figin##1{\vskip2in}\gdef\figins##1{\hskip.5in}%
\else\message{(FIGURES WILL BE INCLUDED)}%
\gdef\figin##1{##1}\gdef\figins##1{##1}\fi}
\def\DefWarn#1{}
\def\figinsert{\goodbreak\midinsert}
\def\ifig#1#2#3{\DefWarn#1\xdef#1{fig.~\the\figno}
\writedef{#1\leftbracket fig.\noexpand~\the\figno}%
\figinsert\figin{\centerline{#3}}\medskip%
\centerline{\vbox{\baselineskip12pt
\advance\hsize by -1truein\noindent\tensl%
\centerline{{\bf Fig.~\the\figno}~#2}}
}\bigskip\endinsert\global\advance\figno by1}
\ifx\answ\bigans
\def\titleft{\titsm}
\magnification=1200\baselineskip=12pt plus 2pt minus 1pt
%
\voffset=0.35truein\hoffset=0.250truein
\hsize=6.0truein\vsize=8.5 truein
\hsbody=\hsize\hstitle=\hsize
\else\let\lr=L
\def\titleft{\titla}
\magnification=1000\baselineskip=14pt plus 2pt minus 1pt
%
\vsize=6.5truein
\hstitle=8truein\hsbody=4.75truein
\fullhsize=10truein\hsize=\hsbody
\fi
\parskip=4pt plus 15pt minus 1pt
\font\titsm=cmr10 scaled\magstep2
\font\titla=cmr10 scaled\magstep3
\font\tenmss=cmss10
\font\absmss=cmss10 scaled\magstep1
\newfam\mssfam
\font\footrm=cmr8  \font\footrms=cmr5
\font\footrmss=cmr5   \font\footi=cmmi8
\font\footis=cmmi5   \font\footiss=cmmi5
\font\footsy=cmsy8   \font\footsys=cmsy5
\font\footsyss=cmsy5   \font\footbf=cmbx8
\font\footmss=cmss8
\def\footfont{\def\rm{\fam0\footrm}
\textfont0=\footrm \scriptfont0=\footrms
\scriptscriptfont0=\footrmss
\textfont1=\footi \scriptfont1=\footis
\scriptscriptfont1=\footiss
\textfont2=\footsy \scriptfont2=\footsys
\scriptscriptfont2=\footsyss
\textfont\itfam=\footi \def\it{\fam\itfam\footi}
\textfont\mssfam=\footmss \def\mss{\fam\mssfam\footmss}
\textfont\bffam=\footbf \def\bf{\fam\bffam\footbf} \rm}
\def\tenpoint{\def\rm{\fam0\tenrm}
\textfont0=\tenrm \scriptfont0=\sevenrm
\scriptscriptfont0=\fiverm
\textfont1=\teni  \scriptfont1=\seveni
\scriptscriptfont1=\fivei
\textfont2=\tensy \scriptfont2=\sevensy
\scriptscriptfont2=\fivesy
\textfont\itfam=\tenit \def\it{\fam\itfam\tenit}
\textfont\mssfam=\tenmss \def\mss{\fam\mssfam\tenmss}
\textfont\bffam=\tenbf \def\bf{\fam\bffam\tenbf} \rm}
\ifx\answ\bigans\def\abstractfont{\tenpoint}\else
\def\abstractfont{\def\rm{\fam0\absrm}
\textfont0=\absrm \scriptfont0=\absrms
\scriptscriptfont0=\absrmss
\textfont1=\absi \scriptfont1=\absis
\scriptscriptfont1=\absiss
\textfont2=\abssy \scriptfont2=\abssys
\scriptscriptfont2=\abssyss
\textfont\itfam=\bigit \def\it{\fam\itfam\bigit}
\textfont\mssfam=\absmss \def\mss{\fam\mssfam\absmss}
\textfont\bffam=\absbf \def\bf{\fam\bffam\absbf}\rm}\fi
%
\def\f@@t{\baselineskip10pt\lineskip0pt\lineskiplimit0pt
\bgroup\aftergroup\@foot\let\next}
\setbox\strutbox=\hbox{\vrule height 8.pt depth 3.5pt width\z@}
\def\vfootnote#1{\insert\footins\bgroup
\baselineskip10pt\footfont
\interlinepenalty=\interfootnotelinepenalty
\floatingpenalty=20000
\splittopskip=\ht\strutbox \boxmaxdepth=\dp\strutbox
\leftskip=24pt \rightskip=\z@skip
\parindent=12pt \parfillskip=0pt plus 1fil
\spaceskip=\z@skip \xspaceskip=\z@skip
\Textindent{$#1$}\footstrut\futurelet\next\fo@t}
\def\Textindent#1{\noindent\llap{#1\enspace}\ignorespaces}
\def\footnote#1{\attach{#1}\vfootnote{#1}}%

\def\foot{\attach\footsymbolgen\vfootnote{\footsymbol}}
\let\footsymbol=\star
\newcount\lastf@@t           \lastf@@t=-1
\newcount\footsymbolcount    \footsymbolcount=0
\def\footsymbolgen{\relax\footsym
\global\lastf@@t=\pageno\footsymbol}
\def\footsym{\ifnum\footsymbolcount<0
\global\footsymbolcount=0\fi
{\iffrontpage \else \advance\lastf@@t by 1 \fi
\ifnum\lastf@@t<\pageno \global\footsymbolcount=0
\else \global\advance\footsymbolcount by 1 \fi }
\ifcase\footsymbolcount \fd@f\star\or
\fd@f\dagger\or \fd@f\ast\or
\fd@f\ddagger\or \fd@f\natural\or
\fd@f\diamond\or \fd@f\bullet\or
\fd@f\nabla\else \fd@f\dagger
\global\footsymbolcount=0 \fi }
\def\fd@f#1{\xdef\footsymbol{#1}}
\def\space@ver#1{\let\@sf=\empty \ifmmode #1\else \ifhmode
\edef\@sf{\spacefactor=\the\spacefactor}
\unskip${}#1$\relax\fi\fi}
\def\attach#1{\space@ver{\strut^{\mkern 2mu #1}}\@sf}
%
\newif\ifnref
\def\rrr#1#2{\relax\ifnref\nref#1{#2}\else\ref#1{#2}\fi}
\def\ldf#1#2{\begingroup\obeylines
\gdef#1{\rrr{#1}{#2}}\endgroup\unskip}

\nreffalse
\def\refout{\footatend\immediate\closeout\rfile\writestoppt
\baselineskip=10pt\newsec{References}{\frenchspacing%
\parindent=12pt\escapechar=` \input
refs.tmp\vfill\eject}\nonfrenchspacing}
%
\def\eqn#1{\xdef #1{(\secsym\the\meqno)}
\writedef{#1\leftbracket#1}%
\global\advance\meqno by1\eqno#1\eqlabeL#1}
\def\eqnalign#1{\xdef #1{(\secsym\the\meqno)}
\writedef{#1\leftbracket#1}%
\global\advance\meqno by1#1\eqlabeL{#1}}
%
\def\chap#1{\newsec{#1}}
\def\chapter#1{\chap{#1}}
\def\sect#1{\subsec{{ #1}}}
\def\section#1{\sect{#1}}
\def\\{\ifnum\lastpenalty=-10000\relax
\else\hfil\penalty-10000\fi\ignorespaces}
\def\note#1{\leavevmode%
\edef\@@marginsf{\spacefactor=\the\spacefactor\relax}%
\ifdraft\strut\vadjust{%
\hbox to0pt{\hskip\hsize%
\ifx\answ\bigans\hskip.1in\else\hskip .1in\fi%
\vbox to0pt{\vskip-\dp
\strutbox\sevenbf\baselineskip=8pt plus 1pt minus 1pt%
\ifx\answ\bigans\hsize=.7in\else\hsize=.35in\fi%
\tolerance=5000 \hbadness=5000%
\leftskip=0pt \rightskip=0pt \everypar={}%
\raggedright\parskip=0pt \parindent=0pt%
\vskip-\ht\strutbox\noindent\strut#1\par%
\vss}\hss}}\fi\@@marginsf\kern-.01cm}
\def\titlepage{%
\frontpagetrue\nopagenumbers\abstractfont%
\hsize=\hstitle\rightline{\vbox{\baselineskip=10pt%
{\abstractfont\pubnum}}}\pageno=0}
\frontpagefalse
\def\pubnum{}
\def\pdate{\number\month/\number\yearltd}
\def\makefootline{\iffrontpage\vskip .27truein
\line{\the\footline}
\vskip -.1truein\leftline{\vbox{\baselineskip=10pt%
{\abstractfont\pdate}}}
\else\vskip.5cm\line{\hss \tenrm $-$ \folio\ $-$ \hss}\fi}
\def\title#1{\vskip .7truecm\titlestyle{\titleft #1}}
\def\titlestyle#1{\par\begingroup \interlinepenalty=9999
\leftskip=0.02\hsize plus 0.23\hsize minus 0.02\hsize
\rightskip=\leftskip \parfillskip=0pt
\hyphenpenalty=9000 \exhyphenpenalty=9000
\tolerance=9999 \pretolerance=9000
\spaceskip=0.333em \xspaceskip=0.5em
\noindent #1\par\endgroup }
\def\autskip{\ifx\answ\bigans\vskip.5truecm\else\vskip.1cm\fi}
\def\author#1{\vskip .7in \centerline{#1}}

\def\address#1{\ifx\answ\bigans\vskip.2truecm
\else\vskip.1cm\fi{\it \centerline{#1}}}
\def\abstract#1{
\vskip .5in\vfil\centerline
{\bf Abstract}\penalty1000
{{\smallskip\ifx\answ\bigans\leftskip 2pc \rightskip 2pc
\else\leftskip 5pc \rightskip 5pc\fi
\noindent\abstractfont \baselineskip=12pt
{#1} \smallskip}}
\penalty-1000}
%

%


\def\bfone{\relax{\rm 1\kern-.35em 1}}
\def\inbar{\vrule height1.5ex width.4pt depth0pt}
\def\IC{\relax\,\hbox{$\inbar\kern-.3em{\mss C}$}}
\def\ID{\relax{\rm I\kern-.18em D}}
\def\IF{\relax{\rm I\kern-.18em F}}
\def\IH{\relax{\rm I\kern-.18em H}}
\def\II{\relax{\rm I\kern-.17em I}}
\def\IN{\relax{\rm I\kern-.18em N}}
\def\IP{\relax{\rm I\kern-.18em P}}
\def\IQ{\relax\,\hbox{$\inbar\kern-.3em{\rm Q}$}}
\def\IR{\relax{\rm I\kern-.18em R}}
\font\cmss=cmss10 \font\cmsss=cmss10 at 7pt
\def\ZZ{\relax\ifmmode\mathchoice
{\hbox{\cmss Z\kern-.4em Z}}{\hbox{\cmss Z\kern-.4em Z}}
{\lower.9pt\hbox{\cmsss Z\kern-.4em Z}}
{\lower1.2pt\hbox{\cmsss Z\kern-.4em Z}}\else{\cmss Z\kern-.4em
Z}\fi}
\def\CP{\relax\ifmmode\mathchoice
{\hbox{\cmss CP}}{\hbox{\cmss CP}}
{\lower.9pt\hbox{\cmsss CP}}
{\lower1.2pt\hbox{\cmsss CP}}\else{\cmss CP}\fi}
\def\a{\alpha}  
 
 \def\l{\lambda}
\def\L{\Lambda}

\def\cH{{\cal H}} 
\def\cJ{{\cal J}} 
\def\cL{{\cal L}} 
 \def\cO{{\cal O}}
 \def\cQ{{\cal Q}}
\def\cR{{\cal R}} 
\def\nup#1({{\it Nucl.\ Phys.}\ $\us {B#1}$\ (}
\def\plt#1({{\it Phys.\ Lett.}\ $\us  {#1}$\ (}
\def\cmp#1({{\it Comm.\ Math.\ Phys.}\ $\us  {#1}$\ (}
\def\prp#1({{\it Phys.\ Rep.}\ $\us  {#1}$\ (}
\def\prl#1({{\it Phys.\ Rev.\ Lett.}\ $\us  {#1}$\ (}
\def\prv#1({{\it Phys.\ Rev.}\ $\us  {#1}$\ (}
\def\mpl#1({{\it Mod.\ Phys.\ Let.}\ $\us  {A#1}$\ (}
\def\ijmp#1({{\it Int.\ J.\ Mod.\ Phys.}\ $\us{A#1}$\ (}
\def\tit#1|{{\it #1},\ }
%

%

\def\ni{\noindent}
\def\tilde{\widetilde}
\def\bar{\overline}
\def\us#1{\bf{#1}}

\def\hat{\widehat}

\def\Coe#1.#2.{{#1\over #2}}
\def\coeff#1#2{\relax{\textstyle {#1 \over #2}}\displaystyle}
\def\coe#1.#2.{\relax{\textstyle {#1 \over #2}}\displaystyle}

\def\shalf{\relax{\textstyle {1 \over 2}}\displaystyle}

\def\to{\rightarrow}
\def\notin{\hbox{{$\in$}\kern-.51em\hbox{/}}}
\def\shdot{\!\cdot\!}

\def\attac#1{\Bigl\vert
{\phantom{X}\atop{{\rm\scriptstyle #1}}\phantom{X}}}

\def\del{\partial}

\def\nex#1{$N\!=\!#1$}

\def\cc{$^,$}

\catcode`\@=12
\def\nul#1{{\it #1}}
\def\brs{$BRST$\ }
\def\qbrs{\cQ_{BRST}}
\def\jbrs{\cJ_{BRST}}
\def\mm#1#2{M_{#1,#2}}

\def\bi#1{b_{#1}}
\def\ci#1{c_{#1}}
\def\zw#1#2.{{#2\over(z-w)^{#1}}}
\def\tg{T_{gh}}
\def\cQ{{\cal Q}}

\def\lv{{Liouville}}
\def\wnpq#1#2#3{M^{(#1)}_{#2,#3}}

\def\sc{superconformal\ }

\def\qt{\tilde\cQ}
\def\zmw#1{(z-w)^{#1}}

\def\g{{\gamma^0}}

\def\LG{Lan\-dau-Ginz\-burg\ }

\def\RG{\cR^{{\rm gr}}}
\def\RN{\cR^{\Atop}}
\def\Atop{A^{{\rm top}}_{k+1}}
\def\mskp{\vskip-.25truecm\noindent}
\def\O{\Omega}
\def\cph#1#2{{\rm CP}_{\!{#1},{#2}}^{\lower3pt\hbox{\fiverm top}}}
\def\cpnk{\cph{n-1}{k}}
\def\rnkg#1#2{\cR^{\cph{#1}{#2}}}
\def\rnk#1{\cR^{\cph{n-1}{#1}}}
\def\wnk#1{W^{\cph{n-1}{#1}}}

\def\L{\Lambda}

\def\Lz{\L^{(z)}}
\def\O{\Omega}
\ldf\mat{D.J.~Gross and A.A.~Migdal, \prl{64} (1990) 717;
M.~Douglas and S.~Shenker, \nup{235} (1990) 635;
E.~Brezin and V.~Kazakov, \plt{236B} (1990) 144.}
\ldf\TOPALG{E.\ Witten, \cmp{117} (1988) 353; \cmp{118} (1988) 411;
\nup340 (1990) 281.}
\ldf\EYtop{T.\ Eguchi and S.\ Yang, \mpl4 (1990) 1693.}
\ldf\Muss{G.\ Mussardo, G.\ Sotkov, M.\ Stanishkov, {\it Int.\ J.\
Mod.\ Phys.}
{\us A4} (1989) 1135; N.\ Ohta and H.\ Suzuki, \nup332(1990) 146.}
\ldf\fusions{M.\ Spiegelglas and S.\ Yankielowicz, \nup393 (1993)
301; D.\ Gepner,\cmp141(1991) 381; M.\ Spiegelglas, \plt274(1992)
21.}
\ldf\MD{M.\ Douglas, \plt238B(1990) 176.}
\ldf\topgr{E.\ Witten,  \nup340 (1990) 281; E.\ and H.\ Verlinde,
\nup348 (1991) 457.}
\ldf\blackh{E.\ Witten,   {\it Phys.\ Rev.} {\us D44} (1991) 314.}
\ldf\TFTmat{R.\ Dijkgraaf and E.\ Witten, \nup342(1990) 486;
R.\ Dijkgraaf and E.\ and H.\ Verlinde, \nup348 (1991) 435;
For a review, see: R.\ Dijkgraaf, \nul{ Intersection theory,
integrable hierarchies and topological field theory}, preprint
IASSNS-HEP-91/91.}
\ldf\Witgr{E.\ Witten, \nup373 (1992) 187. }
\ldf\FLMW{P.\ Fendley, W.\ Lerche, S.\ Mathur and N.P.\ Warner,
\nup348
(1991) 66.}
\ldf\wbrs{M.\ Bershadsky, W.\ Lerche, D.\ Nemeschansky and N.P.\
Warner, \plt B292 (1992) 35; E.\ Bergshoeff, A.\ Sevrin and X.\ Shen,
\plt B296 (1992) 95; J. de Boer and J. Goeree, \nup405 (1993) 669;
E.\ Bergshoeff, H.\ Boonstra, M.\ de Roo S.\ Panda and A.\ Sevrin,
\plt B308 (1993) 34.}
\ldf\DVV{R.\ Dijkgraaf, E. Verlinde and H. Verlinde, \nup{352} (1991)
59.}
\ldf\Loss{A. Lossev, \nul{ Descendants constructed from matter field
and K.
Saito higher residue pairing in Landau-Ginzburg theories coupled to
topological
gravity}, preprint TPI-MINN-92-40-T. }
\ldf\BGS{B.\ Gato-Rivera and A.M.\ Semikhatov, \plt B293 (1992) 72.}
\ldf\LVW{W.\ Lerche, C.\ Vafa and N.P.\ Warner, \nup324 (1989) 427.}
\ldf\cring{D.\ Gepner, \nul{ A comment on the chiral algebras of
quotient
superconformal field theories}, preprint PUPT-1130; S.\ Hosono and
A.\
Tsuchiya, \cmp136(1991) 451.}
\ldf\EXTRA{B.\ Lian and G.\ Zuckerman, \plt254B (1991) 417; P.\
Bouwknegt, J.\ McCarthy and K.\ Pilch, \cmp145(1992) 541; E.\ Witten,
\nup373 (1992) 187.}
\ldf\GG{E.\ Witten, \nup371(1992)191; O.\ Aharony, O.\ Ganor, J.\
Sonnenschein, S.\ Yankielowicz and N.\ Sochen, \nup399 (1993) 527;
O.\ Aharony, J.\ Sonnenschein and S.\ Yankielowicz, \plt B289 (1992)
309; J.\ Sonnenschein, \nul{Physical states in topological coset
models}, preprint TAUP-1999-92; V.\ Sadov, \nul{ on the spectra of
$sl(N)_k/sl(N)_k$ cosets and $W_N$ gravities I}, preprint
HUTP-92/A055.}
\ldf\BMPtop{P.\ Bouwknegt, J.\ McCarthy and K.\ Pilch, \nul{On
physical states
in 2d (topological) gravity}, preprint CERN-TH.6645/92.}
\ldf\Wchiral{K.\ Ito, \plt B259(1991) 73; \nup370(1992) 123; D.\
Nemeschansky and S.\ Yankielowicz, \nul{ N=2 W-algebras,
Kazama-Suzuki models and Drinfeld-Sokolov reduction}, preprint
USC-91-005A;
L.J.\ Romans, \nup369 (1992) 403; W.\ Lerche, D.\ Nemeschansky and
N.P.\ Warner, unpublished.}
\ldf\bo{M.\ Bershadsky and H.\ Ooguri, \cmp126(1989) 49;
M.~Bershadsky and
H.~Ooguri, \plt{229B} (1989) 374.}
\ldf\topw{K.\ Li, \plt B251 (1990) 54, \nup346 (1990) 329; H.\ Lu,
C.N.\ Pope and X.\ Shen, \nup366(1991) 95; S.\ Hosono, \nul{
Algebraic definition of topological W-gravity}, preprint UT-588;
H.\ Kunitomo, Prog.\ Theor.\ Phys.\ 86 (1991) 745.}
\ldf\DK{J.\ Distler and T.\ Kawai, \nup321(1989) 509.}
\ldf\MS{P.\ Mansfield and B.\ Spence, \nup362(1991) 294.}
\ldf\KS{Y.\ Kazama and H.\ Suzuki, \nup321(1989) 232.}
\ldf\Keke{K.\ Li, \nup354(1991) 711; \nup354(1991)725.}
\ldf\Vafa{C.\ Vafa, \mpl6 (1991) 337.}
\ldf\Ind{S.\ Govindarajan, T.\ Jayamaran and V.\ John, \nup402 (1993)
118.}
\ldf\modul{D.\ Kutasov, E.\ Martinec and N.\ Seiberg, \plt B276
(1992) 437.}
\ldf\dis{J.\ Distler, \nup342(1990) 523.}
\ldf\Wstrings{A.\ Bilal and J.\ Gervais, \nup326(1989) 222; P.\
Mansfield and B.\ Spence, \nup362(1991) 294; S. Das, A. Dhar and S.
Kalyana Rama, preprint TIFR/TH/91-20; C.N.\ Pope, L.J.\ Romans and
K.S.\ Stelle, \plt268B(1991) 167; H.\ Lu, C.N.\ Pope, S.\ Schrans and
K.\ Xu, \nup385(1992) 99; C.N.\ Pope, \nul{A Review of W-strings},
preprint CTP-TAMU-30/92; \nul{W Strings 93}, preprint
CTP-TAMU-55-93.}
\ldf\popextra{For extra states at $c_M=100$, see: C.N.\ Pope, E.
Sezgin\ , K.S.
Stelle\ and X.J.\ Wang, \nup299(1993) 247; H. Lu, B.E.W. Nilsson,
C.N. Pope, K.S. Stelle, and P.C. West, Int.\ J.\ Mod.\ Phys.\ A8
(1993) 4071; H. Lu, C.N. Pope, S.\ Schrans and X.\ Wang, \nup
408(1993) 3}
\ldf\BLNWB{M.\ Bershadsky, W.\ Lerche, D.\ Nemeschansky and N.P.\
Warner,
\nup401 (1993) 304.}
\ldf\EYQ{T.\ Eguchi, H.\ Kanno, Y.\ Yamada and S.-K.\ Yang, \plt B305
(1993) 235.}
\ldf\newBMP{P.\ Bouwknegt, J.\ McCarthy and K.\ Pilch, \nul{
Semi-infinite
cohomology of $W$-algebras}, preprint USC-93/11 and ADP-23-200/M15.}
\ldf\gep{D.\ Gepner, \nul{ Foundations of rational quantum field
theory 1},
preprint CALT-68-1825.}
\ldf\MuVa{S.\ Mukhi and C.\ Vafa, \nup407 (1993) 667.}
\ldf\disp{I.\ Krichever, \cmp143 (1992) 415; B.\ Dubrovin, \nup379
(1992) 627, \cmp 145 (1992) 195, \cmp 152 (1993) 539.}
\ldf\Wrev{For reviews on $W$-algebras, see: P.\ Bouwknegt and K.\
Schoutens, Phys.\ Rep.\ 223 (1993) 183; J. \ de Boer, \nul{Extended
conformal symmetry in non-critical string theory}, Ph.D.\ thesis,
1993; J.\ Goeree, \nul{Higher spin extensions of two-dimensional
gravity}, P.D.\ thesis, 1993; T.\ Tjin, \nul{Finite and infinite
$W$-algebras}, P.D.\ thesis, 1993.}
\ldf\PaRoy{S.\ Panda and S.\ Roy, \plt B317 (1993) 533, and preprint
IC-93-307.}
\ldf\DSLG{W.\ Lerche, \nul{Generalized Drinfeld-Sokolov Hierarchies,
Quantum Rings, and W-Gravity}, preprint CERN-TH.6988/93.}
\ldf\flatco{K.\ Saito, J.\ Fac.\ Sci.\ Univ.\ Tokyo Sec.\ IA.28
(1982) 775; M.\ Noumi, Tokyo.\ J.\ Math. 7 (1984) 1; B.\ Blok and A.\
Varchenko, preprint IASSNS-HEP-91/5; E.\ Verlinde and N.P.\ Warner,
\plt 269B (1991) 96; Z. Maassarani, \plt273B (1991) 457; S.\ Cecotti
and C.\ Vafa, \nup367 (1991) 359; W.\ Lerche, D.\ Smit and N.\
Warner,
\nup372 (1992) 87; A.~Klemm, M.~G.~Schmidt and S.~Theisen, \ijmp7
(1992) 6215.}
\ldf\NaSu{T.\ Nakatsu and Y.\ Sugawara, \nup385 (1992) 276.}
\ldf\BBRT{E.\ Bergshoeff, J.\ de Boer, M.\ de Roo and T.\ Tjin,
\nul{On the cohomology of the non-critical W-string}, preprint
UG-7/93.}
\ldf\genDS{M.\ de Groot, T.\ Hollowood and J.\ Miramontes, \cmp145
(1992) 57; N.\ Burroughs, M.\ De Groot, T.\ Hollowood and J.\
Miramontes, \plt B277 (1992) 89; T.\ Hollowood, J.\ Miramontes and
J.\ Guillen, \nul{Generalized integrability and two-dimensional
gravitation}, preprint CERN-TH-6678-92.}
\ldf\DS{Drinfel'd and V.~G.~Sokolov,
 Jour.~Sov.~Math. {\bf 30} (1985) 1975.}
\ldf\krich{I. Krichever, \nul{Topological minimal models and soliton
equations}, Landau Institute preprint.}
\ldf\KosA{B.\ Kostant, Am.\ J.\ Math.\ 81 (1959) 973.}
\ldf\KosB{B.\ Kostant, private communication via N.\ Warner.}
\ldf\fusionr{D.\ Gepner, \cmp 141 (1991) 381.}
\ldf\integr{P.\ Fendley, W.\ Lerche, S.\ Mathur and N.P.\ Warner,
\nup348
(1991) 66; S.\ Cecotti and C.\ Vafa, \nup367(1991)359; D.\
Nemeschansky and N.P.\ Warner, \nup 380(1992) 241.}
\ldf\LW{W.\ Lerche and N.P.\ Warner, \nup358 (1991) 571.}
\ldf\qucoho{E.\ Witten, \cmp 118 (1988) 411, \nup 340 (1990) 281; K.\
Intriligator, \mpl6 (1991) 3543; C.\ Vafa, {\it Topological mirrors
and quantum rings}, in: Essays in Mirror Symmetry, ed.\ S.T.\ Yau,
1992; E.\ Witten, \nul{The Verlinde
algebra and the cohomology of the grassmannian}, preprint
IASSNS-HEP-93/41.}
\ldf\FLMW{P.\ Fendley, W.\ Lerche, S.\ Mathur and N.P.\ Warner,
\nup348
(1991) 66.}
\ldf\getz{E.\ Getzler, \nul{Two-dimensional topological gravity and
equivariant
cohomology}, MIT preprint 1993.}
%
\voffset=0.00truein\hoffset=0.150truein
\hsize=6.0truein\vsize=8.5 truein
%
%
\def\abstr{We review the superconformal properties of matter coupled
to $2d$ gravity, and $W$-extensions thereof. We show in particular
how
the \nex2 structure provides a direct link between certain
matter-gravity systems and matrix models. We also show that much,
probably all, of this can be generalized to $W$-gravity, and this
leads to an infinite class of new exactly solvable systems. These
systems are governed by certain integrable hierarchies, which are
generalizations of the usual KdV hierarchy and whose algebraic
structure is given in terms of quantum cohomology rings of
grassmannians.}

\font\ninerm=cmr9
\font\ninebf=cmbx9
\font\titsm=cmr10 scaled\magstep2
\nopagenumbers
\def\pubnum{
\hbox{CERN-TH.7128/93}
\hbox{hepth@xxx/9401121}}
\def\pdate{
\hbox{CERN-TH.7128/93}
\hbox{December 1993}
}
\titlepage
\vskip 2.5truecm
\title{\titsm Chiral Rings and Integrable Systems\break
for Models of Topological Gravity}
\bigskip
\bigskip
\bigskip
\tenpoint

\font\ninerm=cmr9
\font\ninebf=cmbx9
\centerline{W.\ Lerche}
\bigskip
\centerline{{\it CERN, CH 1211 Geneva 23, Switzerland}}
\bigskip
\vfil
{\centerline{\it Talk given at  }}
{\centerline{\it Strings '93, Berkeley,}}
{\centerline{\it and at}}
{\centerline{\it  XXVII. Internationales Symposium \"uber
Elementarteilchentheorie,}}
{\centerline{\it Wendisch-Rietz}}
\bigskip
\bigskip\vfil
\noindent{\tenrm \abstr}
\vskip 3.truecm
\eject
\def\pdate{}
\centerline{{ \ninebf CHIRAL RINGS AND INTEGRABLE SYSTEMS}}
\centerline{{ \ninebf FOR MODELS OF TOPOLOGICAL GRAVITY}}
\vskip.8truecm
\centerline{{\ninerm W.\ LERCHE}}
\centerline{{\it CERN, CH 1211 Geneva 23, Switzerland}}
\vskip.5truecm
\vbox{\hbox{\centerline{{\ninerm ABSTRACT}}}
{\smallskip\leftskip 3pc \rightskip 3pc \noindent \ninerm
\abstr\smallskip}}

\footline={\hss\tenrm\folio\hss}

\newsec{Introduction}

There has been some recent progress in understanding theories
describing conformal matter coupled to $2d$ gravity.
Matter-plus-gravity systems are interesting to study because they
are, for certain choices of matter theories, supposed to be exactly
solvable. More precisely, they are supposed to be equivalent to
matrix models\mat, which are exactly solvable by themselves as a
consequence of an underlying structure of KdV-type integrable
hierarchies. To deduce this equivalence directly from \lv\ theory
is quite difficult, largely due to technical complications. We
will review how the \nex2 \sc structure helps to provide a manifest
and direct relationship of (certain of) such models to matrix models,
by making use of a connection to topological \LG theory,\Vafa\ and
to integrable hierarchies.

Since theories of $W$-gravity coupled to matter appear by
now\wbrs\cc\newBMP\ to be on a footing similar to ordinary gravity,
one might suspect, by analogy, that there should exist a
corresponding infinite sequence of new types of matrix models that
describe these theories, governed by certain integrable hierarchies.
In a recent paper\DSLG, we made some progress in understanding these
new integrable systems in terms of chiral rings, and we will
briefly review the main ingredients of this construction as well.

\newsec{\nex2 \sc symmetry of the matter-gravity system}

We like to briefly recapitulate ordinary gravity coupled
to conformal $2d$ matter. For simplicity, we will consider mainly
minimal matter models, but this is not really important. These matter
models, denoted by $\mm pq$, where $p,q=1,2,\dots$ are coprime
integers, have central charges $c_M =13 - 6 (t + \coeff1t)$, where
$t\equiv q/p$. We thus consider tensor products
$$
M_{p,q}^{{\rm matter}}\, \otimes\,M_{p,-q}^{{\rm Liouville}}\,
\otimes\,\big\{b,c\big\}\ ,
\eqn\moddef
$$
where $M_{p,-q}^{{\rm Liouville}}$ denotes a Liouville theory with
appropriate central charge, and $\{b,c\}$ denotes the
fermionic ghost system with spins $\{2,-1\}$.

In \brs quantization the physical states of the combined
matter-gravity system are given by the non-trivial cohomology classes
of
a \brs operator,
$$
\qbrs\ =\ \oint\coeff{dz}{2\pi i} \jbrs\ ,
\qquad \ \ \jbrs\ =\ c\,[T_M + T_L +
\shalf T_{gh}] \ ,\eqn\brstcur
$$
which is nilpotent for $c_L+c_M =26$. The most prominent physical
states correspond to the tachyon operators\DK:
$$
T_{r,s}\ = \ c\, V^L_{r;-s}\,V^M_{r;s}\ ,\eqn\tach
$$
where $V_{r;s}$ denotes exponential vertex operators in the usual
notation. By convention, the tachyons have $bc$-ghost number equal to
one. In addition, there exist\EXTRA\ extra physical states whose
number and precise structure depends on the specific value of $t$.
For unitary minimal models, where $t=(p+1)/p$, there exist infinitely
many of such extra states for each matter primary, whereas for
generic $t$, there exists basically only one extra sort of states
besides the tachyons: these are the operators with vanishing ghost
charge. They form what is called\Witgr\ the ground ring, which we
will denote by $\RG$. It is precisely because these operators have
zero ghost charge (and zero dimension like all physical operators),
that the set of ground ring operators closes into itself under
operator products. In fact, even though this ring is in general
infinite, it is finitely generated, ie., it has two generators by
whose action all other ring elements can be generated\Witgr:
$$
\eqalign{
x\ &=\ \big[ bc -
{t\over\sqrt{2t}}(\del\phi_L-i\del\phi_M)\big]\,V^L_{1,2}V^M_{1,2}\cr
\g\ &=\ \big[ bc -
{1\over\sqrt{2t}}(\del\phi_L+i\del\phi_M)\big]\,V^L_{2,1}V^M_{2,1}\
.}
\eqn\gx
$$
(Here, $\phi_{L,M}$ denotes the Liouville field and the matter free
field, respectively). The properties of the ground ring elements
remind very much to the typical features of chiral fields in \nex2
\sc theories. The whole point is, of course, that the
matter-gravity-ghost system is essentially nothing but a (twisted)
\nex2 \sc theory. More precisely, it is known\BGS\cc\BLNWB\ that one
can improve the \brs current \brstcur\ by a total derivative piece,
$$
G^+ ~=~ \jbrs - \del\Big(\sqrt{\coeff2t}  (c \del\phi_L) + \shalf
(1 - \coeff2t) \del c \Big)\ , \eqn\mggplus
$$
such that $G^+$ together with \mskp
$$
G^- \ =\ b\ ,\qquad
T\ ~=~ T_L + T_M + T_{gh}\ ,\qquad
J\ ~=~ c b  + \sqrt{\coeff2t}\,\del \phi_L\ ,
\eqn\currents
$$
indeed generates the (topologically twisted\TOPALG\cc\EYtop) \nex2
\sc algebra,
$$
\eqalign{
T(z)\shdot T(w)\ &\sim\
{2T(w)\over\zmw 2}
+{\partial T(w)\over (z-w)}\ ,\cr
T(z)\shdot G^\pm(w)\ &\sim\
{\coeff12(3\mp 1) G^\pm(w)\over\zmw 2}+{\partial G^\pm(w)\over
(z-w)}\ ,\cr
T(z)\shdot J(w)\ &\sim\ {\coeff 13c^{N=2}\over\zmw 3} +
{J(w)\over\zmw 2}+{\partial J(w)\over (z-w)}\ ,\cr
J(z)\shdot J(w)\ &\sim\ {\coeff 13c^{N=2}\over\zmw 2}\ ,\ \ \ \ \ \
J(z)\shdot G^\pm(w)\ \sim\ \pm {G^\pm(w)\over (z-w)}\ ,\cr
G^+(z)\shdot G^-(w)\ &\sim\
{\coeff 13c^{N=2}\over\zmw 3}+{J(w)\over\zmw 2}+{T(w)+\partial J(w)
\over (z-w)}\ ,\cr
G^\pm(z)\shdot G^\pm(w)\ &\sim\ 0\ ,\cr
}\eqn\tneqtwo
$$
with anomaly \mskp
$$
c^{N=2} ~=~ 3\big(1 - \coeff2t\big) \ . \eqn\minntwo
$$
Upon untwisting, $T\to T-\shalf\del J$, $c^{N=2}$ becomes the central
charge of an ordinary \nex2 algebra. Note that the free-field
realization \mggplus, \currents\ of the \nex2 algebra is different
from the usual one\Muss. This is however irrelevant, and one can
show\BLNWB\ that the above realization can be obtained by hamiltonian
reduction\bo\ from a $SL(2|1)$ WZW model in a way that is analogous
and equivalent to the way of deriving the usual free-field
realization of the \nex2 algebra. Alternatively, one can show that
the two free field realizations of the \nex2 algebra can be
obtained as two different gauge choices in a topological gauged
WZW model\NaSu.

Actually, the construction of the twisted \nex2 algebra is a priori
not unique\MuVa\cc\PaRoy. Indeed, one may replace the \lv\ field
$\phi_L$ in \mggplus, \currents\ by some appropriate combination of
$\phi_L$ with the matter free field, $\phi_M$. However, for
describing minimal models coupled to gravity, the above choices for
$G^+$ and $J$ are the unique, correct ones\foot{Our choice for $J$
implies that it is not holomorphically conserved if the theory is
perturbed by the cosmological constant\MuVa, and one might get the
impression that this is not desirable. For our application to minimal
models, there is however nothing wrong with that.}. Namely, if $G^+$
and $J$ depended on $\phi_M$, then the various different vertex
operator representatives $V^M_{r,s}$ that describe the same given
physical state of the minimal model would have different properties
under the \nex2 algebra, which clearly would not make any sense.
However, for non-minimal models, where these vertex operators
describe distinct physical states, there are other possible choices.
For example, in order to describe black holes in \nex2 language, one
chooses\MuVa\ the \nex2 currents to depend only on the matter field,
$\phi_M$.

An immediate question is about the significance of the twisted \nex2
\sc symmetry. For generic $t$, the mere presence of an \nex2 algebra
doesn't really seem to provide any important new insights, since the
representation theory for arbitrary $c^{N=2}$ is not very
restrictive. On the other hand, for integer $t\equiv k+2$, $k\geq0$,
a lot can be learned: namely then the anomaly \minntwo\ becomes equal
to the anomaly of the twisted \nex2 minimal models, $\Atop$:
$c^{N=2}={3k\over k+2}$. This is a powerful statement, since minimal
models tend to be easily solved entirely by representation theory.
(For $t=-1$, which describes $c\!=\!1$ matter coupled to gravity and
which can be related to the $2d$ black hole\MuVa, the theory is
solvable as well, but we will focus on $t\geq2$ in the following.)

However, this does not yet imply that the minimal models $\mm1{2+k}$
coupled to gravity are the same as the topological minimal models
$\Atop$. What we have shown is simply that these theories have the
same free field realization with the same central charges. A priori,
they don't have even the same spectra. The spectrum of a topological
\nex2 model is well known\EYtop: it is given by the chiral ring\LVW,
which is the finite set of primary chiral fields. For $\Atop$, this
is a nilpotent, polynomial ring generated by one element $x$:
$$
\RN\ =\ {P(x)\over [x^{k+1}\equiv0]}\ =\ \Big\{ 1,x,x^2,\dots,x^k
\Big\}\ .\eqn\nring
$$
One can check that powers of the ground ring generator $x$ in \gx\
are indeed primary and chiral with respect to the \nex2 currents
\mggplus\ and \currents, and that $\RN$ is identical to the subring
of the ground ring $\RG$ that is generated by $x$. (For $t=k+2$, it
turns out that the corresponding tachyons \tach\ have the same \nex2
quantum numbers as the ground ring elements, so that they can be
viewed as different representatives of the same set of physical
fields.) On the other hand, the full ground ring $\RG$ of the
matter-gravity system contains infinitely many more operators\Ind:
$$
\RG\ =\ \RN \otimes \Big\{\,(\g)^n, \ \ n=0,1,2,\dots\,\Big\}\
.\eqn\gring
$$
These extra operators simply do not exist in the topological minimal
models. The difference between the spectra \nring\ and \gring\ can be
accounted for as follows: it turns out that the extra operators are
exact with respect to an additional \brs like operator, $\qt$:
$$
\g =\ -\big\{\qt\,,\,(\coeff{t+1}t\,\del c
+\coeff1{\sqrt{2t}}\,c\,\del\phi_L)\,\big\}\  ,\quad {\rm where}\ \
\qt\ =\
 \oint\coeff{dz}{2\pi i}\,b\,e^{-\coeff
t{\sqrt{2t}}(\phi_L-i\phi_M)}\ .
\eqn\qtilde
$$
One can show\BLNWB\ that $\qt$ is one of the Felder-like screening
operators that arise in our particular free field realization of the
minimal models. That is, {\it by definition} the full \brs operator
of the topologically twisted \nex2 minimal models is the sum of
$\qbrs$, $\qt$ and the other screening operators, such that it
maximally
truncates the infinite free field spectrum precisely to the finite
set of physical operators \nring.

We thus see that the full \brs operator of the topological minimal
models is not the correct one if we wish to describe the minimal
models $\mm1{2+k}$ coupled to gravity. The correct operator obtains
if we drop $\qt$ as an extra piece of the full \brs operator, and it
can be shown that then indeed the ``missing'' operators $(\g)^n$
become physical. This can be actually be better formulated in terms
of equivariant cohomology. Roughly speaking, imposing equivariant
cohomology means that one restricts the Hilbert space to states
$|X\rangle$ that satisfy $b_0|X\rangle=0$. The operator
$(\coeff{t+1}t\,\del c +\coeff1{\sqrt{2t}}\,c\,\del\phi_L)$ in
\qtilde\ does not obey this condition, and this means that the ground
ring generator $\g$ is not the \brs variation of a physical operator
-- hence, it is physical. It is actually
well-known\topgr\cc\BMPtop\cc\getz\
that in pure topological gravity one has to require equivariant
cohomology, in order to obtain a non-empty theory.

The situation can be summarized in Fig.1. What we have
discussed so far corresponds to step A. \ifig\figone{
Models describing gravity coupled to conformal minimal matter of type
$(1,k+2)$.}{\epsfxsize4.5in\epsfbox{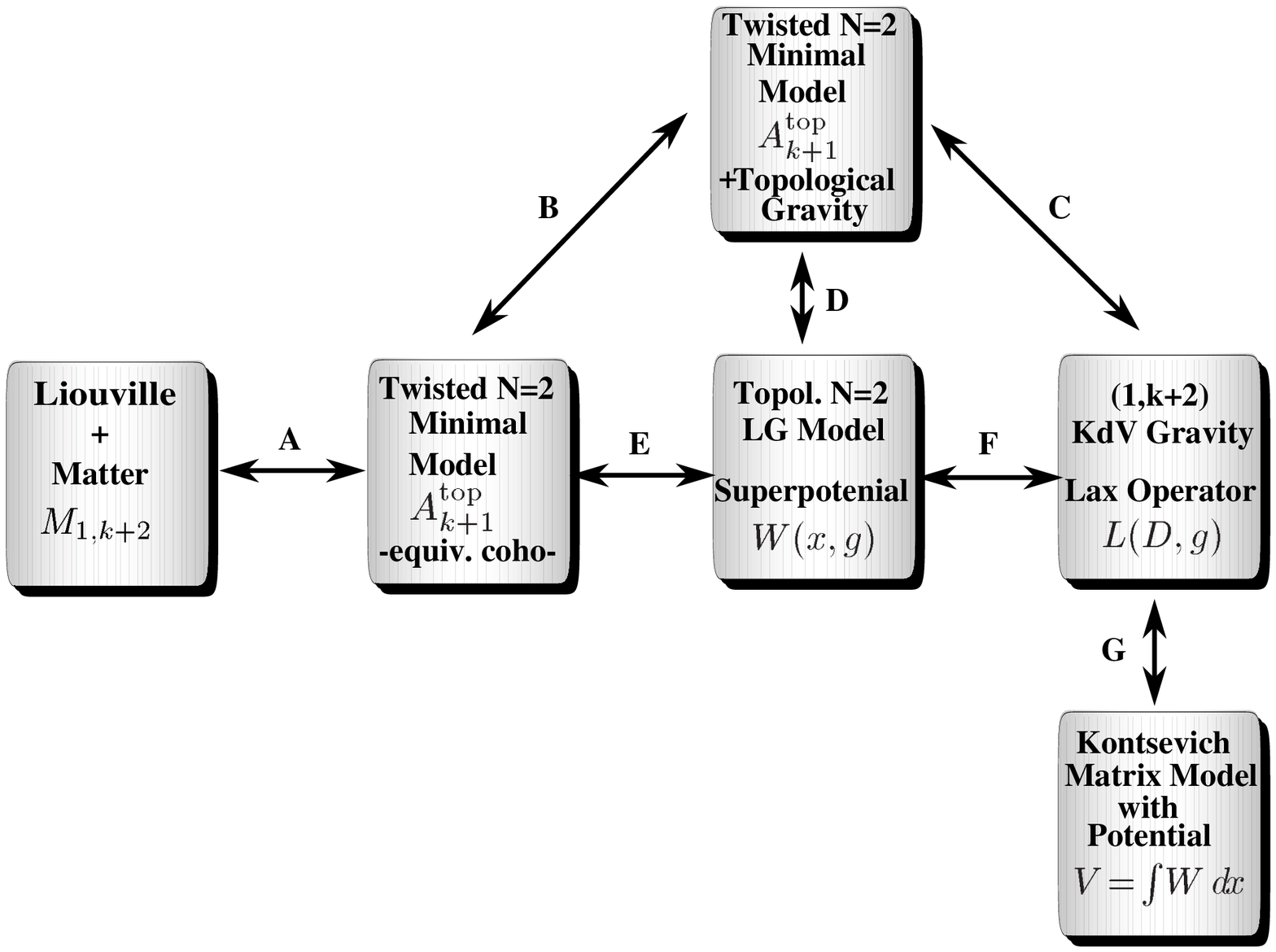}}

It can be shown\EYQ\ that the {\it modified} minimal topological
models, which contain the operators $(\g)^n$ and which are equivalent
to the minimal models $\mm1{2+k}$ coupled to gravity, are in fact
also equivalent to the {\it un-modified} models $\Atop$ coupled to
{\it topological} gravity\topgr. A priori, the building blocks of
$\Atop$ and of the same models coupled\Keke\ to topological gravity
appear to be quite different. There is, however, the remarkable fact
that the total \brs operator of the topological matter plus
topological gravity system obeys\EYQ
$$
\cQ^{tot}\ \equiv \cQ^{N=2} + \qbrs  \ =\ U^{-1}\cQ^{N=2} U \ ,
\eqn\qequiv
$$
where $U=e^{\oint c\,[G_M+G_L+{1\over2} G_{gh}]}$ is a homotopy
operator. The upshot is that the cohomologies of $\Atop$ coupled to
topological gravity and of the modified minimal topological models
are isomorphic, so that at least at the level of Fock spaces the
theories are equivalent. This refers to step B in Fig.1. Step C is an
expression of the fact\Keke\ that the recursion relations of
$[\Atop\otimes{\rm topological\ gravity}]$ are the same as those of
the corresponding matrix models\TFTmat.

\newsec{Relation to dispersionless KdV hierarchy}

The relationship between the matter-gravity system and matrix models
can be exhibited also via a more direct route. One can make use of
the fact that the \LG realization\Vafa\ (step E in Fig.1) of the
topological matter models can be directly related\DVV\ to the KdV
integrable structure of the matrix models (step F). More precisely,
one considers the dependence of correlators on perturbation
parameters $t$, defined by\foot{Note that such perturbations lead
away from the conformal point. We restrict here to the ``small phase
space'', ie., to perturbations generated by the primary fields.}
$\langle \dots e^{\int d^2\!z\,d^2\!\theta\sum_{i=0}^k t_{i+2}
x^{k-i}}\rangle$. It was shown\DVV\ that the effect of such
perturbations can be described in terms of a \LG superpotential of
the form \mskp
$$
W(x,g(t))\ =\ \coeff1{k+2}x^{k+2} -
 \sum_{i=0}^{k} g_{i+2}(t)\,x^{k-i}\ .
\eqn\suppot
$$ \mskp
The coupling constants $g_{i+2}(t_j)$ are certain, in general
non-trivial functions of the perturbation parameters. Since the
correlation functions can easily be computed\DVV\ once one knows
$W(x,g(t))$, solving the theory just amounts to determining these
functions. This can be done by making use of the fact that $t$ are
very particular, namely flat\flatco\ coordinates on the LG
deformation space. Requiring that the appropriate Gau\ss-Manin
connection vanishes, leads to the following differential equations
for $g(t)$:
$$
-\,\del_{t_{i+2}}W(x,g(t))\ =\ \del_x\,\O_{k+1-i}(x,g(t))\ ,
\eqn\dispflo
$$
($i=0,\dots,k$), which involve the hamiltonians
$$
\O_i(x,g(t))\ =\
\coeff1i\,\big((k+2)W\big)^{{i\over k+2}}_+(x,g(t))\ .
\eqn\kdvham
$$
(Here, the subscript ``$+$'' denotes, as usual, the truncation to
non-negative powers of $x$.) The crucial observation\DVV\cc\disp\ is
that under the substitutions $x\to D$ and $W(x,g)\to L(D,g)$, these
equations are nothing but the dispersionless
limit of the KdV flow equations
$$
\del_{t_{i+2}}L(D,g(t)) \ =\ \big[\,(L^{k+1-i\over
k+2})_+,L\,\big](D,g(t))\ .
\eqn\kdv
$$
if one imposes as boundary condition the string equation:
$D\,g_{k+2}=1$. These equations describe\mat\cc\MD (step G in the
figure) the dynamics of the matrix models of type $(1,k+2)$. This
immediately proves the equality of correlation functions (as
functions of the small phase space variables $t$) of the primary
fields with the corresponding correlators of the matrix models (step
F). These arguments, which involve only \nex2 \LG theory, can also
be extended to the gravitational descendants and to some of the
recursion relations they obey\Loss\cc\EYQ (in the small phase space).
In fact, \qequiv\ implies that all states of the matter-gravity
system have \brs representatives in the matter sector alone. That is,
the gravitational descendants can be expressed in terms of the LG
field $x$ (in equivariant cohomology) as well:\EYQ
$$
\sigma_n(\phi_i)(x,t)\ =\ \del_x\,\O_{i+(k+2)n +1}(x,g(t))\
\eqn\gravdesc
$$
(where $\sigma_n(\phi_i)(x,0)\equiv (\g)^n x^i$ in previous
notation).
This is precisely in the spirit of what was said above: the
ingredients of the coupling of $\Atop$ to topological gravity are
already build in the structure of the models $\Atop$ themselves\Loss.
All what is necessary to describe the coupling of these models to
topological gravity is to modify their cohomological definition. The
fact that topological gravity coupled to $\Atop$ can be described
purely in terms of LG theory corresponds to step D in Fig.1.

Of particular interest is the perturbation of these models by the
``cosmological constant'' term. In our language\BLNWB\cc\MuVa, it is
the perturbation by the top element of $\RN$,
$$
S_{{\rm cosm}}=\mu \int d^2\!z\, e^{\sqrt{{2\over t}} \phi_L}\
\equiv\ t_2 \int d^2\!z\,d^2\!\theta\, x^k\ .
\eqn\cosmo
$$
It is known\integr\ that this perturbation is integrable and leads to
the massive quantum \nex2 sine-Gordon model; although not invariant
under the full (twisted) \nex2\goodbreak\noindent \sc symmetry, it is
supersymmetric, and the corrected supercharge $\oint G^+$ still
serves as a \brs operator. Under the perturbation, the \nex2 $U(1)$
current $J$ ceases to be holomorphically conserved, which is a
typical feature of perturbations leading away from the conformal
point. The effective superpotential is given by a Chebyshev
polynomial:
$$
W(x,t_2=\mu)\ =\
\coeff2{k+2}\mu^{{k+2}\over2}\,T_{k+2}(\shalf\mu^{-{1\over2}}x)\ =\
\coeff1{k+2}\,x^{k+2}-\mu\, x^k + O(\mu^2)\ .
\eqn\cheby
$$
At $\mu=1$, the deformed chiral ring becomes identical\fusions\ to
the fusion ring of the $SU(2)_k$ WZW model, which it is also the same
as the
operator product algebra of the $SU(2)_k/SU(2)_k$ topological field
theory.
This observation then allows to finally make contact to the
formulation of
matter-plus-gravity models in terms of topological $G/G$ theories\GG:
it is known\GG\cc\MuVa\ that at the level of Fock space cohomology,
the (suitably defined) $SU(2)/SU(2)$ model is indeed equivalent to
the
matter-gravity system. We thus have the relation:
$$
\Big[\,\mm1{2+k}\otimes{\rm \lv\
gravity}\,\Big]\attac{\mu=1}\!\!\cong\ \
\Big[\,{SU(2)_k\over SU(2)_k}\attac{{{\rm modified}\atop{\rm
cohomology}}}\!\!\!\!\!\Big]
\eqn\otherequivs
$$

\newsec{Extension to $W$-gravity}

One obvious motivation for investigating generalizations is
the wish to step beyond the $c_M\!=\!1$ barrier of ordinary gravity.
This can be achieved by considering matter theories with extended
symmetries,
coupled to the corresponding extended geometry.
The prime candidates for such models are those related to
$W$-algebras. (One may also consider supersymmetric versions:
it turns out\BLNWB\ that \nex1 matter coupled to \nex1 supergravity
yields
\nex3 superconformal models, etc.).

For a given theory of $W_n$-gravity coupled to matter, there is a
barrier at $c_M\!=\!n\!-\!1$, below of which there is a finite number
of (dressed) primary fields and below of which the theory should be
solvable. In analogy to ordinary gravity, one would expect that
such theories should be solvable also at the accumulation points
$c_M\!=\!n\!-\!1$ (where there exists an extra $SU(n)$ current
algebra symmetry). At these points, such models are presumably
related to black hole type of objects in spacetimes with signature
$(\!n\!-\!1,\!n\!-\!1)$ and are characterized by topological field
theories based on non-compact versions of $\cpnk$.

\ni The physical models in question are tensor products
$$
 W_n^{\rm matter}\otimes W_n^{\rm \lv}\otimes_{j=1}^{n-1}
\{b_j, c_j\} \ , \eqn\wstring
$$
which might be called ``non-critical $W$-strings''\Wstrings. Above,
$W_n^{\rm matter}$ denotes conformal field theories that have a
$W$-algebra as their chiral algebra, which can be for example
$W_n$ minimal models $\wnpq npq$ with central charges $c_M = (n-1)[
1-n (n+1) {(t-1)^2\over t}],\ t\equiv q/p$. Furthermore, $W_n^{\rm
\lv}$
denotes a $(n-1)$-component generalization of \lv\ theory (Toda
theory), and $\{b_j, c_j\}$ denotes the Hilbert space of a ghost
system with spins $j+1$ and $-j$, respectively. As it turns out, the
structure of these theories for arbitrary $n$ is very similar to
$n=2$, which corresponds to ordinary gravity. However, only for $n=3$
the generalization\foot{The existence of \brs currents for arbitrary
$n$ can be inferred from indirect arguments\BLNWB\cc\wbrs.} of the
\brs current is explicitly known\wbrs:
$$
\eqalign{
\jbrs\ &=\ \ci2 \big[\coeff1{b_{L}}W_L  + \coeff i{b_{M}} W_M \big]
+ \ci1 \big[T_L +T_M +\shalf\tg^{1} +\tg^{2} \big]\cr
&+ \big[T_L -T_M \big]\bi1 \ci2(\del\ci2)
+ \mu(\del\bi1)\ci2(\del^2\!\ci2)+ \nu\bi1 \ci2(\del^3\!\ci2)\
,}\eqn\jb
$$
where $b^2_{L,M}\equiv{16\over5c_{L,M}+22}$ and $\mu={3 \over 5}\nu=
{1 \over
10{b_L}^2} (1-17{b_L}^2)$. In this equation, $T_{L,M}$ and $W_{L,M}$
denote the
usual stress tensors and $W$-generators of the \lv\ and matter
sectors, and
$\tg^i$ are the stress tensors of the ghosts.

Using this \brs current, one can study the spectrum of physical
operators of $W_3$ matter coupled to $W_3$ gravity, and one
finds\BLNWB\cc\popextra\cc\newBMP\cc\BBRT\ that the analogs of ground
ring
elements and tachyons are states with ghost numbers equal to
$0,1,2,3$ (the first number corresponds to ground ring elements, and
the last one to tachyons). The explicit expressions are however too
complicated to be written down here.

The interesting point is that there appears an \nex2 \sc
symmetry for all $n$. For example, for $W_3$ gravity one finds that
$$
\eqalign{ G^+\ &=\ \jbrs +
\del\Big[\,
-c_{1}J+2i\sqrt{\coeff t3} b_{1} c_{1} c_{2}J +
  i\coeff{\left( 1 + t \right) }{2} \sqrt{\coeff3t}
   b_{1} c_{1} (\del c_{2})
   \cr&-
    i\coeff{\left( 3 + 2 t \right) }{\sqrt{3 t}}
   b_{1} (\del c_{1}) c_{2} -
    \coeff{(7 {t^2}-10t-15)}{4 t}   b_{1} (\del^{2}\!c_{2}) c_{2}+
  i\coeff{(t-9)}{\sqrt{3 t}} b_{2} (\del c_{2}) c_{2}
  \cr&-
  i\coeff{\left( 3 + 4 t \right) }{\sqrt{3 t}} (\del b_{1}) c_{1}
c_{2}-
  \coeff{3 (4 {t^2}-2t-3)}{2 t} (\del b_{1}) (\del c_{2}) c_{2} +
 \coeff{(t-3)}{t} (\del c_{1})
  \cr&+
 i\coeff{1}{2 \sqrt{3 t}} c_{2}[2t {J^2}-3 (t-5)T_L-3 (t-1)T_M
 -6( 1 + t )\del J]
  \cr&+
  i\coeff{ \left( 1 + t \right) }{2} \sqrt{\coeff3t}(\del c_{2})J-
  i\coeff{({t^2}-4t-1)}{2t}\sqrt{\coeff3t} (\del^{2}\!c_{2}) +
  t b_{1} (\del c_{2}) c_{2}J
  \, \Big]\ ,}\eqn\wbrs
 $$
together with \mskp
$$
\eqalign{
G^- \ &=\ b_1\ ,\qquad\ \ \ \
T\ ~=~ T_L + T_M + T_{gh} \ ,\cr
J\ ~&=~ c_{1} b_{1}  + c_{2} b_{2}  +
  \coeff3{\sqrt t}(\l_1\cdot \del \phi_L)
 +\coeff i2\sqrt{\coeff3t}  (t-1) \del[b_{1}c_{2}] \cr
}\eqn\wcurrents
$$
gives a non-standard free field realization of the topological
algebra \tneqtwo\ with
$$
c^{N=2} ~=~ 6\big(1 - \coeff3t\big) \ . \eqn\cks
$$
Since we are dealing here with theories with an extended symmetry,
coupled to an extended ``$W$-geometry'', it is perhaps not too
surprising to find that these topological algebras actually extend to
topologically twisted \nex2 $W$-algebras. For $t=n+k$, which
corresponds to $W_n$-minimal matter models $\wnpq n1{n+k}$, the
anomaly indeed becomes equal to the central charges of the minimal
\nex2 $W_n$ models at level $k$: $c^{N=2}=3{(n-1)k\over n+k}$. These
models are just the well-known Kazama-Suzuki models\KS\ based on
cosets $SU(n)_k\over U(n-1)$, which are known to have an \nex2 $W_n$
chiral algebra\Wchiral. The models that arise here are of course the
topologically twisted versions, which we will denote by $\cpnk$;
$n=2$ corresponds to ordinary gravity coupled to matter:
${\rm CP}_{\!1,k}^{{\rm top}}\equiv\Atop$.

The chiral rings of these topological minimal $W_n$ matter models are
well understood\LVW\cc\cring, and are described further below. They
are generated by primary chiral fields $x_i$, $i=1,\dots,(n-1)$ (with
$U(1)$ charges equal to $i/(n+k)$), and have elements
$$
\cR^{\cpnk}\ =\
\Big\{\prod_{i=1}^{n-1}(x_i)^{n_i}\ ,\sum n_i\leq k\,\Big\}\ .
\eqn\ksrings
$$ \mskp
The full ground rings of the minimal models $\wnpq n1{n+k}$ coupled
to $W_n$-gravity contain in addition generators $\gamma^0_i$,
$i=1,\dots,(n-1)$ (with $U(1)$ charges equal to $i$) and are the
``$W$-gravitationally extended'' chiral rings of the Kazama-Suzuki
models:
$$
\RG\ =\ \cR^{\cpnk} \otimes
\Big\{\prod_{i=1}^{n-1}(\gamma^0_i)^{n_i}\ ,\
n_i=0,1,2,\dots\,\Big\}\ .\eqn\fullring
$$ \mskp
These rings have an obvious interpretation in terms of topological
minimal $W_n$-matter $\cpnk$ coupled to topological
$W_n$-gravity\topw. Like for ordinary gravity, the ground ring
generators $x_i$ can be interpreted as the fields of topological LG
models, with superpotentials given in refs.\LVW\cc\fusionr\ and in
eq.\ (5.20) below. It would be very interesting to investigate as to
what extent also the $W$-gravitational descendants $(\gamma^0_i)$ can
be expressed in terms of LG fields. Ideally, the whole topological
$W_n$-matter-gravity system can be described in terms of \LG theory.

Although this has not yet been thoroughly investigated for general
$n$, our considerations seem so far to indicate that the structure
for general $n$ is indeed very much parallel to the one of $n\!=\!2$.
Accordingly, one would have for $W_n$-matter models of type
$(1,k\!+\!n)$ coupled to $W_n$-gravity a scheme that is analogous to
Fig.1. It would be exciting to verify the remaining links in the
figure for $W$-gravity. In particular, by analogy to step F in Fig.1
one would expect the existence of an infinite sequence of new
integrable systems, whose Lax operators are given in terms of the
Kazama-Suzuki superpotentials, and in analogy to step G one would
expect the existence of an infinite class of new matrix models. While
this latter assertion is more difficult to prove, we were so far
indeed successful\DSLG\ to get an idea about the structure of the new
integrable systems, and this is what we like to briefly outline next.
\goodbreak

\newsec{Quantum rings and integrable systems for $W$-gravity}

The issue is to find a multi-variable generalization\DSLG\ of the
dispersionless KdV hierarchy, which describes the models $\wnpq
n1{n+k}$ coupled to $W_n$-gravity, as well as the models $\cpnk$
coupled to topological $W_n$-gravity. Our strategy is inspired by a
general relationship between topological LG theory, chiral rings and
Drinfeld-Sokolov types of integrable systems. At the heart of our
construction is the generalization of the above-mentioned
relationship between LG superpotential and dispersionless Lax
operator to many variables $x_i$.

We like first to reformulate the ordinary,
dispersionless\disp\ KdV hierarchy\foot{ With ``KdV
hierarchy'' we will always mean the $(k\!+\!1)$th generalized KdV
hierarchy.} (pertaining to the LG models ${\rm CP}_{\!1,k}^{{\rm
top}}\equiv\Atop$) in matrix language, because it is this form of the
hierarchy that is most suitable for our generalization. One starts
with the linear Drinfeld-Sokolov system\DS
$$
\Big[\,D\bfone - \cL_1\,\Big]\cdot \Psi\ =\ 0\ ,
\eqn\firstorder
$$
where the ``Lax operator'' $\cL_1$ is given by the $(k+2)\times(k+2)$
dimensional matrix
$$
\cL_1(g)\ =\  \Lz_1 + Q_1(g)\ ,
\eqn\Ldef
$$
where $\Lz_1$ has the familiar form
$$
\Lz_1\  =\ \pmatrix
{ 0 & 1 & 0 & \dots & 0 \cr
  0 & 0 & 1 & \dots & 0 \cr
  \vdots & \vdots & \vdots & \ddots &  \vdots \cr
0 & 0 & 0 & \dots & 1 \cr
z & 0 & 0 & \dots & 0 \cr}\ \ ,
\eqn\zstep
$$
with $z$ representing the spectral parameter. In \Ldef, $Q_1$ is
usually taken to be a lower triangular matrix that is determined only
up to gauge transformations belonging to the nilpotent subgroup
$N^-$. Upon recursively solving for the components of $\Psi$ in favor
to the first component $\Psi_0$, the system \firstorder\ is
equivalent to the gauge invariant, scalar spectral equation
$$
L(D,g)\,\Psi_0\ =\ \coeff1{k+2}\,z\,\Psi_0\ .
\eqn\scalspec
$$
In the dispersionless limit, where $D\to x$ and $L(D,g)\to W(x,g)$
(cf., \suppot), this is precisely the characteristic equation of the
Lax operator $\cL_1$, which therefore must satisfy
$$
W(\cL_1(g),g)\ =\ \coeff1{k+2}\,z\,\bfone\ .
\eqn\wident
$$
This ``superpotential spectral equation'' can be taken as the
definition of $\cL_1$ in terms of the \LG superpotential $W$, and
(non-uniquely) determines $Q_1(g)$. The gauge freedom can be fixed by
going to any particular gauge. The choice that is most appropriate
for us is however not given by taking $Q_1$, as usual, to be a lower
triangular matrix, but by taking $Q_1$ to belong\genDS\ to the
Heisenberg subalgebra generated by $\Lz_1$ ($Q_1$ is then lower
triangular only up to $\cO(1/z)$). That is, we have an infinite
expansion
$$
\cL_1(g)\ =\ \Lz_1 + \sum_{l=1}^\infty q_l(g) (\Lz_1)^{-l}\ ,
\eqn\Lexpan
$$
whose coefficients can be computed from \wident\ in a recursive way.

\ni The KdV flow equations \dispflo\ that determine
the LG couplings $g(t)$ take the form
$$
\del_{t_{i+2}}\,\O_{k+1-j}(g(t))\ =\
\del_{t_{j+2}}\,\O_{k+1-i}(g(t))\ ,\eqn\matrixflows
$$
and involve the following, matrix-valued hamiltonians:
$$
\eqalign{
 &\O_i(\cL_1(g),g)\ =\
\coeff1i(\Lz_1(\cL_1(g),\cL_1^{-1}(g),g))^i_+\ ,\cr &{\rm with}\
\big[\,\O_i\,,\,\O_j\,\big]\ \equiv0\ ,\qquad\ \ \O_1\equiv\cL_1\ ,}
\eqn\matrixhams
$$
where the subscript ``+'' denotes the truncation to positive powers
of $\cL_1$ in the expansion of the constant matrix $(\Lz_1)^i$. It is
clear that the constant flows associated with
$\O_{n(k+2)}=\coeff1{n(k+2)}\,z^n\bfone$ are trivial and correspond
to perturbations by the null operators $\sigma_n(\phi_{k+1})$; the
hamiltonians $\O_i$ with $i>k\!+\!1$ correspond to the gravitational
descendants (cf., \gravdesc).

It is well-known\DS\cc\genDS\ that the basic underlying structure of
the KdV integrable system is the algebra generated by $\Lz_1$, which
is the principal Heisenberg subalgebra of $\hat{s\ell}(k\!+\!2)$. Its
positive part,
$$
\cH^+\ \equiv\ \big\{\,(\Lz_1)^m, m\in\ZZ_+\,\big\}\ ,
\eqn\heis
$$
is precisely what determines the hamiltonians, $\O=(\cH^+)_+$.
In view of our later generalization, it is very helpful
to note that $\Lz_1$ is identical to the chiral ring structure
constant $C_1(z)$ that pertains to the following LG potential
``at one level higher'':
$$
W^{A^{{\rm top}}_{k+2}}(x,t_{k+2}=z,t_l=0)\ =\
\coeff1{k+3}x^{k+3}- z\, x\ .
\eqn\higherW
$$
This means that the underlying algebraic structure of the $A_{k+1}$
type matter-gravity system is that of a specifically deformed chiral
ring pertaining to the LG theory $A_{k+2}$:
$$
\cH^+\ \cong\ \cR^{A^{{\rm top}}_{k+2}}(t_{k+2}\!=\!z,t_l\!=\!0)\ .
\eqn\HRrel
$$
For $g=0$, the superpotential spectral equation \wident\
represents a specific relation in this ring, and can be viewed as
the equation of motion associated with the LG potential \higherW:
$$
\eqalign{
W^{A^{{\rm top}}_{k+1}}(x,0) - \coeff1{k+2}z\ &=\
\coeff1{k+2}\,\del_x\,
W^{A^{{\rm top}}_{k+2}}(x,z)\cr
&=\ 0\ .
}\eqn\wrel
$$
This important fact, namely that $\Lz_1=C_1(z)$ so that the
matrix-valued spectral equation
$W(\Lz_1)\equiv\coeff1{k+2}(\Lz_1)^{k+2}=\coeff1{k+2} z \bfone$ can
be interpreted as some chiral ring vanishing relation (associated to
a {\it different} LG theory), is our starting point of the
generalization to many variables. More precisely, our plan is to use
appropriate ring structure constants $C_i(z)$ to construct
hamiltonians and Lax operators for the models $\cpnk$ coupled to
gravity. This is motivated by the fact that their chiral rings have a
common underlying structure for all $n$: it is the structure of
principal embeddings\KosA\ of $s\ell(2)$. Such kind of embeddings is
also precisely what underlies the construction of the
$W_n$-algebras.\Wrev

Specifically, it is well-known that the Heisenberg algebra generator
$\Lz_1$ that figures in the Drinfeld-Sokolov matrix
system\DS\cc\genDS\ is nothing but an $s\ell(2)$ step generator $I_+$
(principally embedded in $s\ell(k\!+\!2)$),
$$
\eqalign{
\Lz_1\ &=\ \L_1 + z\, \L_{-(k+1)}\ ,\qquad {\rm where}\cr &\cr
\L_1\ \ \ &=\ I_+\ \equiv\ \sum_{{\rm{simple\atop roots\ \alpha}} }\,
E_\alpha\ , \ \qquad
\L_{-(k+1)}\ =\ E_{-\psi}\ ,\cr
}\eqn\prinSL
$$
perturbed by the spectral parameter $z$ ($\psi$ denotes the highest
root). The point is, as mentioned above, that this is also the
structure of the perturbed chiral ring $\rnkg1{k+1}(z)$: it is
known\LVW\cc\cring\ that the unperturbed ring $\rnkg1{k+1}$ is
isomorphic to the cohomology ring $H^*({\rm CP}_{k+1})$, and there is
a theorem by Kostant\KosB\ that says that $H^*({\rm CP}_{k+1})$ is
generated by an $s\ell(2)$ step generator $I_+$. The deformation by
the spectral parameter is then precisely what deforms the cohomology
ring $H^*$ into the quantum cohomology ring $QH^*$, whence
$$
\cH^+\ \cong\ \rnkg1{k+1}(z)\ \cong\
QH^*_{\bar \del}\big({\rm CP}_{k+1},\IR\big)\ .
\eqn\qcohR
$$
(The word ``quantum'' indicates that the deformation of the classical
cohomology ring by the spectral parameter $z$ is precisely
the effect of the instanton corrections in a supersymmetric
${\rm CP}_{k+1}$ $\sigma$-model\qucoho).

For the more general models $\cpnk$ that are related to
$W_n$-gravity, the story is very similar: it is known\LVW\cc\cring\
that the chiral rings are isomorphic to the Dolbeault cohomology
rings of certain grassmannians:
$$
\rnk k\ \cong\ H^*_{\bar\del}\big(\,\coeff{SU(n+k-1)}{SU(n-1)\times
SU(k)\times U(1)}\,,\IR\,\big)\ .
\eqn\grasscoho
$$
These chiral rings are generated by ring structure constants $C_i$,
$i=1,\dots,(n\!-\!1)$, which represent the LG fields $x_i$. The
important point is that these ring structure constants are determined
by principal embeddings of $s\ell(2)$ as well !

\ni More precisely, consider the following matrices:
$$
\L_p\ =\ \sum_{\{\a:\rho_G\cdot\a=p\}}a_\a^{(p)} E_\a\ ,\ \ \
{\rm\  for\ each\ } p\in\{1,2,\dots,(n+k-2)\}\ ,
\eqn\Fpdef
$$
where the coefficients $a_\a^{(p)}$ are determined trough
$[\L_p,\L_{p'}]=0$, with $\L_1\equiv I_+$ as in \prinSL. Kostant's
theorem\KosB\ now tells that when taken in the $(n\!-\!1)$th
fundamental representation of $s\ell(n\!+\!k\!-\!1)$, the matrices
$\L_i$, $i=1,\dots,(n\!-\!1)$, generate
$H^*(\coeff{SU(n+k-1)}{SU(n-1)\times SU(k)\times U(1)})$, and this
means that they are precisely the ring structure constants of the
Kazama-Suzuki models:
$$
C_i\ =\ \L_i\ .
\eqn\CLrel
$$
Our idea is to employ the $\L_i$ to construct Lax operators and
hamiltonians for generalized Drinfeld-Sokolov systems. The relevant
objects are of course matrices $\Lz_i$ that are perturbed by a
spectral parameter; they are uniquely defined by requiring
$[\Lz_i,\Lz_{i'}]=0$ where $\Lz_1$ is as in \prinSL\ (but now in the
$(n\!-\!1)$th fundamental representation of $s\ell(n\!+\!k\!-\!1)$).
They generate precisely the quantum deformation\qucoho\ of the
grassmannian cohomology rings, which are the same as specifically
perturbed chiral rings of the models $\cpnk$:
$$
\rnk k(t_{k+n-1}=z)\ \cong\ QH^*_{\bar\del}
\big(\,\coeff{SU(n+k-1)}{SU(n-1)\times
SU(k)\times U(1)}\,,\IR\,\big)\ .
\eqn\RQH
$$
These perturbed chiral rings are associated with the LG
superpotentials
$$
\wnk k(x_i,z)\ =\ \wnk k(x_i,0) - z\, x_1\ ,
\eqn\pertW
$$
which were investigated previously\LW\ in the context of integrable
perturbations of the models $\cpnk$. Such superpotentials were
first explicitly written down in refs.\LVW\cc\FLMW, and have the
form:
$$
\eqalign{
\wnk k(x_i,0)\ &=\ \sum_{l=1}^k\,(\xi_l)^{n+k}(x_i)\ ,\cr {\rm
where}\ \qquad\ \ x_i\ &=\!\!\sum_{1\leq l_1\leq\dots\leq l_i\leq
k}\!\!\! \xi_{l_1}\xi_{l_2}\dots\xi_{l_k}
}\eqn\Wunpertdef
$$
are the elementary symmetric polynomials. This formula was obtained
by making use of the fact that, in the Borel-Weil picture, the
cohomology of the grassmannian $G/H$ is generated by Chern classes
$c_i$ of certain $H$-valued vector bundles, which satisfy
relations of the form
$$
{\rm Ch}_{(n-1,1)}(c_i,t)\,\cdot\,{\rm Ch}_{(1,k)}(c_i,t)\ =\ 1\ ,
\eqn\crels
$$
where
$$
{\rm Ch}_{v}(c_i,t)\ =\ \sum_{j=0}^{{\rm dim}\,v} c_j(\xi)\,t^j
\eqn\chern
$$
is the total graded Chern form associated with the $H$-representation
$v$. The relations among the $c_i\cong x_i$ generated by \crels\ lead
precisely to the vanishing relations associated with the potentials
\Wunpertdef. The formula \Wunpertdef\ for the superpotentials was
subsequently used in in ref.\fusionr, where the following
generating function was found:
$$
-\log\Big[\sum_{i=1}^{n-1}(-t)^ix_i\,\Big]\ =\
\sum_{k=-n+1}^{\infty}t^{n+k}\,\wnk k(x_i,0)\ .
\eqn\genallW
$$
{}From this it is easy to prove that
$$
\wnk k(x_i,0)\ =\ \coeff1{n+k}\,
\Big(\sum_{i=1}^{n-1}(n-i)\,x_{i-1}\del_{x_i}\Big)\,\wnk {k+1}(x_i,0)
\ ,\eqn\key
$$
which means that a given superpotential can be written as a vanishing
relation of the superpotential ``at one level higher''. This is the
key point which makes the whole construction fly. Namely,
\pertW\ and \key\ imply that
$$
\wnk k(x_i,0) - \coeff1{n+k}\,z\ =\ \coeff1{n+k}\,
\Big(\sum_{i=1}^{n-1}(n-i)\,x_{i-1}\del_{x_i}\Big)\ \wnk
{k+1}(x_i,z)\ ,
\eqn\fly
$$
and this means that the ring structure constants $\Lz_i=C_i$ of the
models $\cph{n-1}{k+1}$ satisfy $\wnk
k(\Lz_i,0)=\coeff1{n+k}\,z\,\bfone$. This is precisely what we have
been looking for: namely we can take for the Lax operators of the
integrable hierarchies just the perturbed versions of these $\Lz_i$,
$$
\cL_i(g)\ =\ \Lz_i + Q_i(g)\ \equiv\ \Lz_i + \sum_{l_j}
q_{l_1,\dots,l_{n-1}}^i(g)\,
(\Lz_1)^{-l_1}\dots(\Lz_{n-1})^{-l_{n-1}}\ ,
\eqn\lexpan
$$
whose coefficients $q(g)$ are such that
$$
\wnk k(\cL_1(g),\dots,\cL_{n-1}(g),g)\ =\ \coeff1{n+k}\,z\,\bfone\ .
\eqn\desired
$$
This is the desired generalization of the matrix-valued
superpotential spectral equation \wident.

\ni To obtain a hierarchy of differential equations, we need to
construct appropriate hamiltonians. By analogy to \matrixhams, we
simply take the commuting matrices
$$
\O_{l_1,\dots,l_{n-1}}(\cL_i,g)\ =\
\big(\,(\Lz_1)^{l_1}\dots(\Lz_{n-1})^{l_{n-1}}\big)_+\ ,\qquad\ \
l_i\geq0\ ,
\eqn\newhams
$$
where ``+'' denotes projection to positive grade. That is, we take
as relevant Heisenberg algebra
$$
\cH^+\ \cong\ QH^*_{\bar\del}
\big(\,\coeff{SU(n+k)}{SU(n-1)\times
SU(k+1)\times U(1)}\,,\IR\,\big)\ ,
\eqn\Hplus
$$
which just means, like previously for $n=2$, {\it that the underlying
algebraic structure of the $\cpnk$ matter-gravity integrable system
is given by the quantum ring associated with the matter model ``at
one level higher'', $\cph{n-1}{k+1}$}. The perturbation by the
spectral parameter $z$ deforms the finite, nilpotent ring $\rnk{k+1}$
into an infinite dimensional, affine algebra, which reflects the
extension of the matter ring \ksrings\ to the gravitationally
extended ground ring \fullring\ of the matter-gravity system. It
would be very interesting to study in more detail the structure
$\cH^+$ in relation with the $W$-gravity descendants of \fullring.
How this precisely works is not so clear because the number of
hamiltonians \newhams\ per grade does not grow indefinitely with
increasing grade, since there are relations between polynomials of
the $\Lz_i$ (for example the superpotential spectral equation). These
relations are just the multi-generator analogs of the well-known
condition that reduces the KP to the KdV hierarchy.

Strictly speaking, $\cH^+$ in \Hplus\ is the enveloping algebra of
the principal Heisenberg algebra. That is, since the generators
$\Lz_i$ are in general in a higher fundamental representation of
$s\ell(n\!+\!k)$, powers of the $\Lz_i$ will in general not belong to
the principal Heisenberg subalgebra of $\hat{s\ell}(n\!+\!k)$, but to
its enveloping algebra. This is precisely how this construction makes
it possible to have more commuting hamiltonians at a given grade as
compared to the usual KdV type of systems\DS\cc\genDS, where one
considers only the flows associated with the algebra, which are
representation-independent.

The flow equations that determine the small phase space couplings
$g(t)$ have then supposedly the generic form
$$
\eqalign{
\big[\,&D_{l_1,\dots,l_{n-1}},D_{l_1',\dots,l_{n-1}'}\,\big]\ =\ 0\
,\cr &D_{l_1,\dots,l_{n-1}}\ \equiv\ {\del\over \del
t_{l_1,\dots,l_{n-1}}}\, -\,
\sum_{k_j}Z_{l_1,\dots,l_{n-1}}^{k_1,\dots,
k_{n-1}}\,\O_{k_1,\dots,k_{n-1}}(\cL_i(g(t)),g(t))
}\eqn\floeqs
$$
(where $Z$ are normalization constants),
but whether these equations really determine the correct LG couplings
$g(t)$ in terms of the flat coordinates $t$, is a problem that we
don't know how to answer yet in general. All what we have done so far
was to check these equations for a couple of examples, where they
indeed produced the correct results\DSLG.

But these results as well as the general structure strongly suggest
that that the kind of integrable systems we proposed makes sense and
correctly describes the quasi-classical dynamics of the models
$M^{(n)}_{1,n+k}$ coupled to $W_n$-gravity, which are supposedly
equivalent to the models $\cpnk$ coupled to topological
$W_n$-gravity.
This would correspond to the completion of step F in Fig.1 for
$W$-gravity, and make step G in the figure appear feasible.

\newsec{Acknowledgements}
\vskip-.2truecm
I would like to thank the organizers of the conference for their hard
work, and for providing an opportunity for me to present this
material.
\vskip.2truecm
\refout
\end